\documentclass[twocolumn,pra,aps,showpacs]{revtex4}

\usepackage{psfrag,graphicx}

\usepackage{dcolumn}
\usepackage{amsmath,amssymb}
\usepackage{bm}  
\usepackage{pstricks}

\def\half{\textstyle\frac{1}{2}}

 \newcommand{\ket}[1]{|#1\rangle}
 \newcommand{\bra}[1]{\langle #1|}

\begin{document}


\title[Short Title]{Twisted Order Parameter applied to Dimerized Ladders}

\author{
J. Almeida$^{\star}$, M.A. Martin-Delgado$^{\star}$ and G.
Sierra$^{\ast}$
 }
\affiliation{ $^{\star}$Departamento de F\'{\i}sica Te\'orica I,
Universidad Complutense. 28040 Madrid, Spain.
\\
$^{\ast}$Instituto de F\'{\i}sica Te\'orica, C.S.I.C.- U.A.M.,
Madrid, Spain. }

\begin{abstract}
We apply the twisted order parameter (TOP) for dimerized 
quantum spin ladders to locate the critical phases
that separate gapped phases representing quantum spin liquids
of various types. Using the DMRG, method we find that
the TOP is a good order parameter for these systems
regardless the number of legs.
As a check, we reproduce with DMRG and periodic boundary conditions 
the computations previously done with Quantum Montecarlo
for one-dimensional S=1/2, S=1, S=3/2 and S=2 Heisenberg chains with 
alternating bonds.
\end{abstract}


\maketitle

\section{Introduction}
\label{sect_intro}

The physics of quantum spin ladders with dimerization \cite{snake_ladders96}
is an emblematic example of quantum phase transtions with
a rich structure of both gapless and gapped phases \cite{affleck_haldane87,affleck88,strongly}.
The main issue in their quantum phase diagrams is to determine the location of
critical points or lines, which typically separate massive phases. These phases 
exhibit quite strongly correlated properties like Haldane phases \cite{haldane82}, 
an example of quantum spin liquid of great interest in condensed matter systems.
Analytical methods have been applied to uncover this rich phase structure, some of
them non-perturbative 
\cite{snake_ladders96, kotov_et_al99, wang_nersesyan00, chitov1, chitov2}. 
However, they are not sufficient to study with enough precision the whole range
of coupling parameters entering in their quantum lattice Hamiltonians and, consequently,
they have to be complemented with numerical studies
\cite{lanczos_2leg98, okamoto03, FAF2legladder, FAF3legladder}, 
which sometimes modify the conclusions
achieved with analytical methods, either quantitative and/or qualitatively.

Quantum spin ladder systems are not mere theoretical constructs, but they have
been experimentally synthezised in several types of materials
\cite{dagotto_rice96, dagotto, hiroi_et_al91, batlogg_et_al95, ferro_experiment}, 
some of them as interesting as the superconducting cuprates compounds or 
another classes of materials which allow for not only antiferromagnetic couplings
but also for ferromagnetic ones as we shall be considering in this work.
Ladder systems are also valuable candidates to implement their physics in 
simulable optical lattices \cite{greiner,OLs}, 
where the coupling constants of the models could
be externally manipulated more easily than with standard materials.
Furthermore, quantum spin chains and ladders have played a central role
for ground test numerical simulations when the density matrix renormalization group
(DMRG) was introduced \cite{white92, white93, hallberg06, scholl05, dmrg_book}.

A diversity of variants of quantum spin liquids appear as gapped phases in these
dimerized spin ladders. These types of spin orders have been initially classified
resorting to the string order parameter (SOP) \cite{sop1, sop2}.
However, this initial SOP parameter was only valid for systems with spin magnitude
$S=1$. Then, a generalization valid for both integer and half-interger spin systems was
introduced \cite{oshikawa92} and recently, we have been able to check their
validity in a large class of dimerized spin ladders \cite{FAF3legladder} and
we have found that the generalized SOP serves as a good order parameter to distinguish
the many quantum phases appearing in those ladders when their number of legs 
and spin couplings are varied.

When moving from one gapped phase to another, this process generically implies crossing
a critical point or line. Although the SOP parameter can detect the change associated
to this phase crossing, however it does not perform so well when trying to locate
the position of those critical points. To this end, another order paramter called
twisted order parameter (TOP) has been introduced for spin chains 
\cite{TOP_Nakamura_Todo}
and some ladders \cite{TOP_Nakamura_Todo2, matsumoto_et_al03}. 
The TOP being non-local, it can be considered as a close relative of the SOP.
In fact, we shall see that they play complementary roles: the SOP is better suited
for classifying quantum spin liquids, whereas the TOP is more appropriate for
finding their critical points.

The idea behind the TOP comes from the twist operator introduced in the proof
of the Lieb-Schulz-Mattis theorem (LSM) \cite{LSM, affleck_lieb86} in one-dimensional
quantum spin systems. Under certain circumstances, namely, for half-interger $S$-spin
chains, the system is proved to be gapless by creating a sort of twisted excited
state along the chain, whose energy gap with respect to the ground state vanishes
in the thermodynamic limit as $1/N$, where $N$ is the length of the chain.
Although this twist operator cannot be conclusive for the case of integer $S$-spin
chains, thereby opening the door for the celebrated Haldane conjecture \cite{haldane82},
it turns out that it can be used to locate the critical points of dimerized chains
of arbitrary spins, either integer of half-integer. The way to see this connection
is by means of the valence bond solid (VBS) states used as trial wavefunctions to
qualitatively represent those gapped quantum phases.

The existence of extensions of the LSM theorem for spin ladders \cite{affleck88}
can also be used as a hint to trying to extend the notion of the TOP parameter
to more complicated lattices beyond one-dimensional systems.
For spin ladders with $S=\half$ spins, this generalized LSM theorem only works
for an odd number of legs. Interestingly enough, we have found that the generalized
TOP parameter works for locating the critical points, regardless of the number of legs.

This paper is organized as follows:
in Sect.\ref{sect_II} we introduce a set of 
quantum lattice Hamiltonians for spin ladders which present two
different patterns of dimerization, either staggered or
columnar, as well as antiferromagnetic or ferromagnetic couplings
between their legs (chains);
in Sect.\ref{sect_III} we present our numerical results for the TOP
parameter measured in those dimerized spin ladders using DMRG calculations.
With this information we can plot the TOP vs. the dimerization strength
and detect critical points by the vanishing of the TOP parameter. 
Upon varying the dimerization patterns and the ferro- or antiferro-types of
rung couplings, the number of zeros and shape of the TOP also changes
in characteristic forms that serve us to distinguish among the set of
quantum ladder Hamiltonians introduced in the previous section.
Sect.\ref{sect_conclusions} is devoted to conclusions.

\section{Models of dimerized spin ladders}
\label{sect_II}

In this paper we study quantum spin ladders that
can be viewed as a certain number of bond-alternated Heisenberg chains
stacked one on top of another and coupled between them via a
coupling constant  $J'$ corresponding to a Heiserberg-like
spin interaction.
The bond alternation between neighbor spins on the same leg is such that
every strong bond is followed by a weaker one. Every leg can then begin
either with a strong or weak coupling, and that initial choice determines the
dimerization pattern all the way through the ladder.

Then, to characterize completely our ladders we have  to specify
some parameters:
the number of legs $n_l$, the ferromagnetic $J'<0$ or antiferromagnetic $J'>0$
nature  of the Heisenberg coupling between legs and the
dimerization pattern  of every leg, given by the coupling constant $\gamma$.

The Hamiltonian that gives raise to this general set of families of
dimerized Heisenberg ladders is:
\begin{equation}                                                                
H^{\textrm{OBC}}= \sum_{\ell} \sum_{i}J_{i,\ell}\mathbf{S}_{i}(\ell) \cdot      
\mathbf{S}_{i+1}(\ell)                                                          
+J'\sum_{\ell,i} \mathbf{S}_{i}(\ell) \cdot                                     
\mathbf{S}_{i}(\ell+1)                                                          
\label{HamOBC}                                                                  
\end{equation}
where $\mathbf{S}_{i}(\ell)$ denotes a $S=1/2$ spin operator at the site $i$
in the leg $\ell$, with $i=1\dots N$ and $\ell=1\dots n_l$. The set of couplings
denoted by $J_{i,\ell}$ endows the ladders with certain dimerization 
patterns to be specified below.

We shall focus our attention on two and three leg ladders, in addtition to
spin chains. These cases will suffice to clarify the behavior of the 
TOP parameter under diverse dimerization patterns. Moreover,
among the different possibilities of establishing a dimerization pattern on every leg, we
will only be interested in two particular arrangements of bonds. We will denote
the \textit{staggering} dimerization pattern that with a $J_{i,\ell}$ distribution of the form
$J_{i,\ell}=1+(-1)^{i+\ell}\gamma$ while the \textit{columnar} pattern has a distribution
$J_{i,\ell}=1-(-1)^{i}\gamma$, i.e., all the legs begin with a strong bond. In both cases
the parameter $\gamma$ is constrained to be within $-1\le\gamma\le1$.

We will also be interested in using not only open boundary conditions but also periodic ones.
In this case, we have to modify the Hamiltonian \eqref{HamOBC}
to take into account the interactions between both ends:
\begin{equation}
\label{HamPBC}
H^{\textrm{PBC}}=H^{\textrm{OBC}}+
\sum_{\ell}J_{N,\ell}\mathbf{S}_{N}(\ell) \cdot
\mathbf{S}_{1}(\ell)
\end{equation}

Eq. \eqref{HamOBC} is not translationally invariant. On the contrary, if we consider
ladders with an even number(greater than two) of sites per leg and using 
periodic boundary conditions, the Hamiltonian \eqref{HamPBC} is  invariant
under translations of the form
 $\textbf{\textrm{S}}_i(\ell)\rightarrow\textbf{\textrm{S}}_{i+2}(\ell)$ and
 also under reversal
 of the dimerization strength $\gamma\leftrightarrow -\gamma$.

\begin{figure}
\includegraphics[width=6cm]{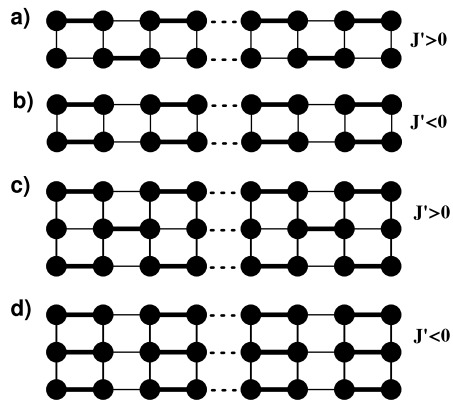}
\caption{Different dimerization patterns in two and three legs spin ladders:
 a) and c) are staggered, b) and d) are columnar. Thicker lines 
correspond to a coupling constant 
$J_{i,\ell}=1+\gamma$ and thiner ones to $J_{i,\ell}=1-\gamma$}
\label{ladderpics}
\end{figure}

The number of possible combinations of those ingredients described above is still high, 
and not all of them may have critical properties.
This fact will depend essentially on the staggering pattern and coupling constant $J'$.
In particular, for ladders with two and three legs, the models that may have critical 
phases for certain configurations of the microscopic parameters 
are the following ones:
i/ the staggered antiferromagnetic two leg ladder,
ii/ the columnar ferromagnetic two leg ladder, 
iii/ the staggered antiferromagnetic three leg ladder and 
iv/ the columnar ferromagnetic three leg
ladder. These models are depicted in fig. \ref{ladderpics}.

\section{Results}
\label{sect_III}

Much like the string order parameter (SOP), the TOP is a non-local parameter
since it involves non-local measurements.
However, unlike the SOP,  the success of the TOP relies 
in the $2\pi$ twist carried out between both ends of the system. This fact
makes it unclear how will the DMRG perform when measuring it. 
This is in sharp contrast with the SOP case,
which can be acurately measured using
only a representative subsystem within the bulk of the whole system.
\begin{figure}
\begin{center}
\includegraphics[width=8.5cm]{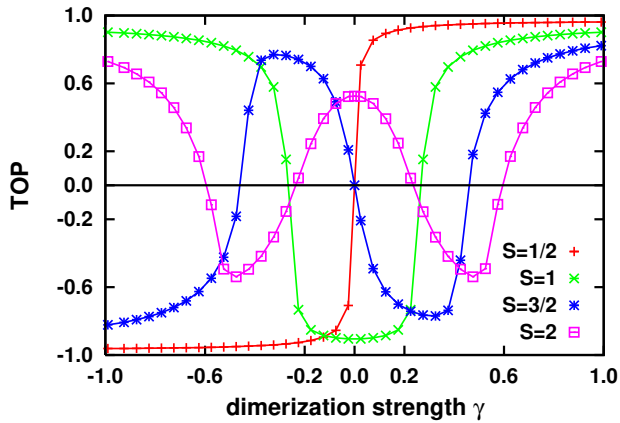}
\caption{ DMRG computation of TOP for different one dimensional Heisenberg chains with 
alternated bonds. The size of the chains is $L=64$ and periodic boundary conditions
have been used.} 
\label{TOPchains}
\end{center}
\end{figure}

\subsection{Spin Chains}
\begin{figure}
\begin{center}
\begin{tabular}{cc}
a)&\includegraphics[width=5.8cm]{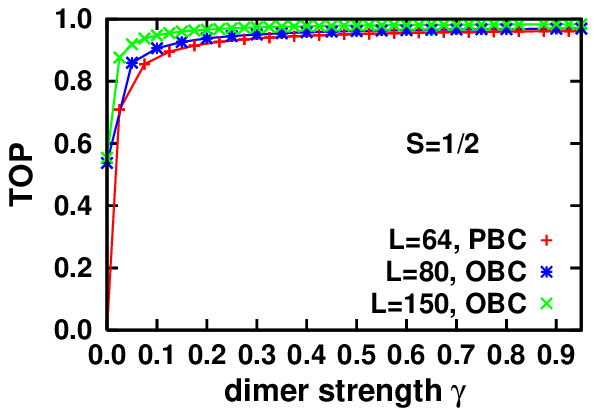}\\
b)&\includegraphics[width=5.8cm]{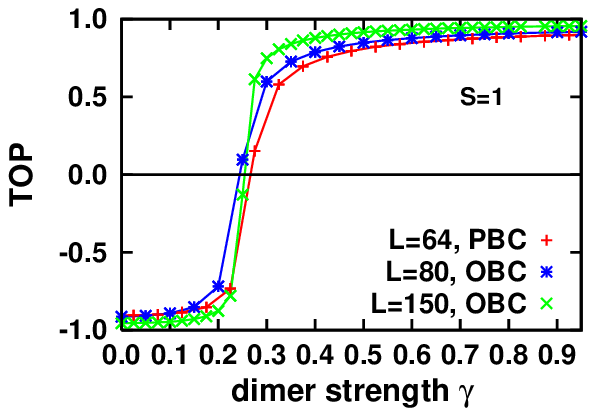}\\
c)&\includegraphics[width=5.8cm]{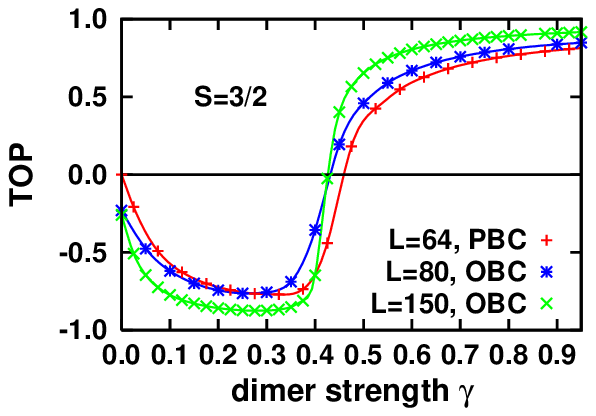}\\
d)&\includegraphics[width=5.8cm]{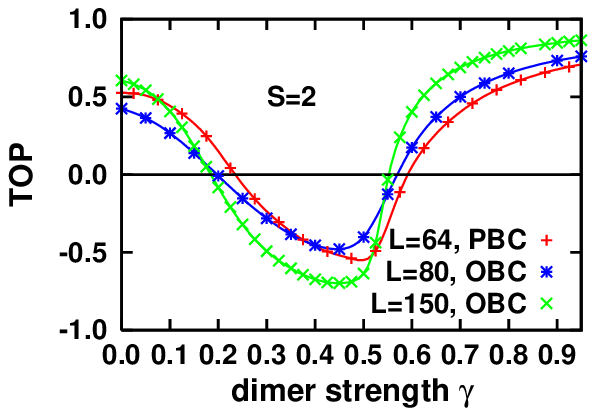}\\
\end{tabular}
\caption{
TOP computed in various Heisenberg chains with alternated
bonds and different spin values: a) $S=1/2$, b) $S=1$, c) $S=3/2$, d) $S=2$.
Computations using open boundary conditions properly capture the location 
of the critical points as we increase the size of the system, however 
the convergence of the TOP towards zero at the particular point 
$\gamma_c=0$ in the graphs a) and c) is very slow and makes a difference
between both types of boundary conditions.
}
\label{TOPsubfigs}
\end{center}
\end{figure}

Due to this uncertainty and as a first
check, we have reproduced the Quantum Montecarlo results for one dimensional dimerized chains 
\cite{TOP_Nakamura_Todo} using DMRG-adapted methods. Fig. \ref{TOPchains}
 shows the TOP computed for these
chains using periodic boundary conditions. To compute these values the number of
DMRG sweeps was set to three, the number of retained states of the density matrix was 
$m=200$ and the Lanczos tolerance equal to $10^{-10}$. With these parameters, the
results obtained using DMRG and Quantum Montecarlo agree almost identically. In order to
go a bit further and obtain better estimates of the critical points using DMRG, 
it is more convenient to use open ladders. 
Fig. \ref{TOPsubfigs} represents the TOP 
computed using open boundary conditions and different sizes of the chains. 
The largest size
 $L=150$ was computed performing four DMRG sweeps, $m=300$ and the tolerance equal to 
$10^{-9}$, the less demanding case $L=80$ required two sweeps, $m=120$ and the tolerance 
equal to $10^{-8}$.
 The critical points of each chain are those where the TOP is zero. Looking at the different
graphs two observations are in order: on one hand we see that the location of the critical
points is overestimated in the case of periodic boundary conditions compared to the case of 
open chains. On the other hand, the values of the critical points computed with 
open boundary conditions are in good agreement with the values of these points computed in
the thermodinamic limit in other works \cite{TOP_Nakamura_Todo}.  In fact, from fig. 
\ref{TOPsubfigs} and the curves with $L=150$ we see that for the $S=1$ chain we obtain
 $\gamma_c=0.254$, for $S=3/2$ $\gamma_c=0.426$. The case with $S=2$ has two critical 
points at $\gamma_c=0.184$ and $\gamma_c=0.522$. These values can be respectively compared
with the previously computed $\gamma_c=0.25997(3)$,
$\gamma_c=0.43131(7)$,$\gamma_c=0.1866(7)$,$0.5500(1)$.

Regarding the critical points $\gamma_c=0$ in cases with $S=1/2$ and $S=3/2$, we 
observe that the TOP unambiguously goes to zero using periodic boundary conditions. 
With open boundary conditions, however, the TOP is still far from zero even with 
$L=150$, although the tendency is clearly to vanish as we increase the size.
Interestingly enough,  a similar slow converging behaviour
 precisely at $\gamma_c=0$ has been
also observed in \cite{FAF3legladder} measuring the string order parameter.
  
\subsection{Spin Ladders}
In the rest of this section we will measure the TOP in the models described 
previously in Sect.~\ref{sect_II} 
in order to test wether or not it can also be succesfully applied 
to non strictly one
dimensional cases.

 We commented in the case of the spin chains that the TOP using open boundary
conditions converged very slowly to zero at $\gamma_c=0$ in those models that in fact 
possesed such a critical point. Since some of the ladders that we will study in this section
have indeed a critical point close to $\gamma=0$ we will use periodic boundary conditions to
avoid the slow convergence of these points. Moreover, as we commented before, with periodic 
boundary conditions our models are invariant under sign reversal of the dimerization strength.
 This means that for a given computed critical point, say $\gamma_c$, there must exist another
one equal to $-\gamma_c$. The difference in the value of these opposite points will then
 give us a 
hint about the truncation error of our DMRG computations.

The natural extension of the TOP as defined in \cite{TOP_Nakamura_Todo} to systems with many legs
is as follows:
\begin{equation}
z_\ell^1:=\bra{\psi_0} \textrm{exp}\Big\{\frac{2\pi i}{N}\sum_{j=1}^N jS^z_j\Big\}\ket{\psi_0}
\end{equation}
Where each operator now includes the contribution coming from each leg of the ladder
$S_i=\sum_{\ell}S_i(\ell)$.\\

\begin{figure}
\begin{center}
\includegraphics[width=8.0cm]{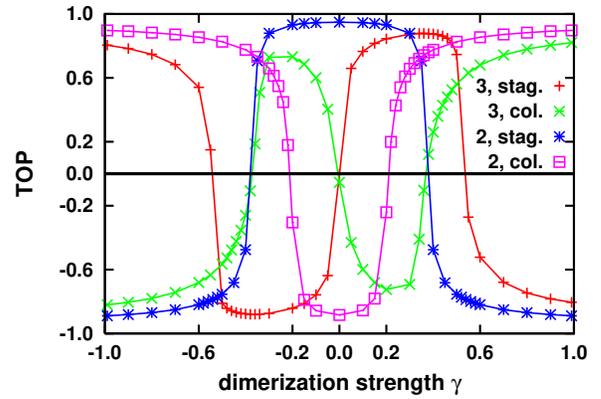}
\caption{ TOP computed in ladders with columnar and staggering dimerization.
The constant $J'$ has a fixed value equal to $J'=-5$ in the two and three leg 
ladder with columnar dimerization and $J'=1$ in the two and three leg with staggered 
dimerization. Boundary
conditions have been used. The size of the two leg ladders is $L=2\times60$, and $L=3\times60$ 
for the three leg ladders}
\label{TOPladders}
\end{center}
\end{figure}

In fig. \ref{TOPladders} we have computed the TOP on each one of the critical ladders described
in the previous section. To run the DMRG we have set $m=340$, the tolerance equal to
$10^{-9}$ and performed three sweeps.
For each one of the ladders we have used an arbitrary value of
the constant $J'$.
 The critical lines of these models have been obtained by other techniques in
references \cite{lanczos_2leg98,FAF2legladder,FAF3legladder}. Since we are using 
periodic boundary conditions we are constrained to use smaller sizes than those considerered
in the references and hence no exact agreement is expected. With this consideration in mind,
the location of the critical points in fig. \ref{TOPladders}, that is, 
the points where the TOP is zero, is 
in good agreement with those of the references, as comes out from table \ref{TOPtable}.\\

\begin{table}[htb]
\begin{tabular}{llll}
&&&\\
Ladder&\hskip1ex $J'$&\hskip12ex TOP&\hskip5ex Other\\
\hline
&&&\\
3, stagg.&  1.0 & $\gamma_c=0.001, -0.542, 0.536$ 
& $\gamma_c=0, \pm 0.527$\\ 
3, col.&  -5.0 & $\gamma_c=0.005, -0.372, 0.368$
& $\gamma_c=0, \pm 0.335$\\  
2, stagg.&  1.0 & $\gamma_c=-0.38, 0.38$
& $\gamma_c=\pm 0.34$\\  
2, col.&  -5.0 & $\gamma_c=-0.212, 0.211$ 
& $\gamma_c=\pm 0.196$\\  
\end{tabular}\\
\caption{ 
Location of some critical points of different dimerized ladders. The identification of
each ladder is done by means of the number of legs: two or three,
 and the dimerization pattern: staggered
or columnar. We provide
the values obtained with the TOP and the corresponding ones computed
using different techniques. References related to these last values are 
given in the text.
}
\label{TOPtable}
\end{table}

The sign of the TOP is related with the nature of the VBS phase. From its definition 
it can be seen that its sign is given by a factor $(-1)^k$, where $k$ is the number of
 isolated spins at the end of the chain. More explicitly, in a general $(m,n)$-VBS state, $k=m$ 
if the number of sites is odd and $k=n$ if it is even.\\
We see in fig. \ref{TOPladders} that the sign of the TOP is coherent with the VBS phases
of each ladder. That is, in the cases of the two and three leg ladders with columnar
dimerization the TOP is negative from $\gamma=0$ up to the corresponding critical values
$\gamma_c$ where the TOP is zero, and then the TOP is positive all the way to $\gamma=1$. 
On the other
hand the behavior of the TOP with the two and three leg ladders with staggered dimerization is
the contrary. Since our computations have been all done using an even number of sites per leg,
the sign of the TOP agrees with the characterization of the phases of each model done
by means of the generalized SOP in refs. \cite{FAF2legladder,FAF3legladder}.
 That is, moving from $\gamma=0$ to $\gamma=1$
the two leg ladder with columnar dimerization moves from a $(1,1)$-VBS to a $(2,0)$-VBS while
the three leg ladder with columnar dimerization moves from a $(2,1)$-VBS to a $(3,0)$-VBS.
In these cases $k=1$ and the sign of the TOP is negative in the first phases, and $k=0$ and
the sign of the TOP is positive in the second phases. In the case of the three leg ladder 
with staggered dimerization we move from a $(1,2)$-VBS to a $(2,1)$-VBS and again the sign
of the TOP is correct according to the value of $k$. We have no information on the VBS phases
of the two leg ladder with staggered dimerization but in this particular case the only
possibility to obtain a negative value of the TOP is a $(1,1)$-VBS, while a positive TOP
could be a $(2,0)$-VBS or a $(0,2)$-VBS.

\section{Conclusions}
\label{sect_conclusions}

The results shown in the previous section reveal that the DMRG algorithm is 
well suited to compute the twisted order parameter even with open boundary conditions.
 We have applied TOP parameter
to four different spin ladder systems and estimated the location of their critical points.
We find agreement with other DMRG calculations which compute the vanishing of the gap
directly. We have checked that the TOP computation
is not only restricted to strictly one dimensional systems but it also works in two- and 
three-leg ladders. This holds true despite the LSM theorem based on the twist operator
only works for ladders with an odd number of legs and $S=\half$ spins.
In this sense, the twisted order parameter serves as a suitable and complementary tool
to the string order parameter in the characterization of quantum phases in dimerized
spin systems.

\bigskip
\noindent {\em Acknowledgements}:
Part of the computations of this work were performed with the High
Capacity Computational Cluster for Physics of UCM (HC3PHYS UCM), funded
in part by UCM and in part with FEDER funds.
We acknowledge financial support from  DGS grants  under contracts BFM 2003-05316-C02-01,
FIS2006-04885, and the ESF Science Programme INSTANS 2005-2010.


\begin{thebibliography}{99}

\bibitem{snake_ladders96}
M. A. Martin-Delgado, R. Shankar and G. Sierra,
``Phase Transitions in Staggered Spin Ladders'';
Phys. Rev. Lett. {\bf 77}, 3443 (1996).

\bibitem{affleck_haldane87}
 I. Affleck, F. D. M. Haldane,
"Critical theory of quantum spin chains";
Phys. Rev. {\bf B}  36 (1987) 5291.
\bibitem{affleck88}
I. Affleck, Les Houches 1988-Session XLIX: {\em Fields, strings
and critical phenomena} (North-Holland, Amsterdam 1990), chap.
Field Theory Methods and Quantum Critical Phenomena, pp. 563-640.


\bibitem{strongly}
{\em Strongly Correlated Magnetic and Superconducting Systems},
Lecture Notes in Physics, Proceedings of the El Escorial Summer School 1996.
Eds. G. Sierra and M. A. Martin-Delgado
Springer-Verlag (1997).

\bibitem{haldane82}
F. D. M.  Haldane,
"Continuum dynamics of the 1-D Heisenberg antiferromagnet:
 Identification with the O(3) nonlinear sigma model",
Phys. Lett. {\bf A} 93, 464-468 (1982).


\bibitem{kotov_et_al99}
V. N. Kotov, J. Oitmaa  and Zheng Weihong;
"Excitation spectrum and ground-state properties of the S=1/2 Heisenberg ladder with staggered dimerization"
Phys. Rev. B {\bf 59}, 11377 - 11383 (1999).

\bibitem{wang_nersesyan00}
Y.-J. Wang and A. A. Nersesyan,
"Ising model description of the SU$(2)_1$ quantum critical point in a dimerized two-leg spin-1/2 ladder"
Nucl. Phys. {\bf B} 583, 671 (2000).


\bibitem{chitov1}
M. Azzouz, K. Shahin, G.Y. Chitov,
``Spin-Peierls instability in the spin- 1/2 Heisenberg three-leg ladder'';
Phys. Rev. \textbf{B} 76, 132410 (2007).

\bibitem{chitov2}
G.Y. Chitov, B. Ramakko, M. Azzouz;
``Quantum Criticality in Dimerized Spin Ladders'',
arXiv0709.3256C.



\bibitem{lanczos_2leg98}
M. A. Martin-Delgado, J. Dukelsky and G. Sierra,
"Phase diagram of the 2-leg Heisenberg ladder with alternating dimerization".
Phys. Lett. {\bf A} 250, 87 (1998).

\bibitem{okamoto03}
K. Okamoto,
"Phase diagram of the S=1/2 two-leg spin ladder with staggered bond alternation"
Phys. Rev. {\bf B} 67, 212408 (2003).

\bibitem{FAF2legladder}
J. Almeida, M.A. Martin-Delgado, G. Sierra,
"DMRG study of the Bond Alternating \textbf{S}=1/2 
Heisenberg ladder with Ferro-Antiferromagnetic couplings". 
Phys. Rev. B \textbf{76}, 184428 (2007).


\bibitem{FAF3legladder}
J. Almeida, M.A. Martin-Delgado, G. Sierra,
"Critical lineas and massive phases in quantum spin ladders with
dimerization''; arXiv:0707.4452 (2007)



\bibitem{dagotto_rice96}
E. Dagotto, T. M. Rice,
"Surprises on the Way from One- to Two-Dimensional Quantum Magnets: The Ladder Materials",
Science {\bf 271} 618 - 623, (1996).


\bibitem{dagotto}
E. Dagotto, Rep. Prog. Phys. \textbf{62} 1525-1571.
"Experiments on ladders reveal a complex interplay between a spin-gapped normal state and superconductivity",

\bibitem{hiroi_et_al91}
Z. Hiroi, M. Azuma, M. Takano and Y. Bando, J. Sol. State Chem. 95, 230 (1991).

\bibitem{azuma_et_al94}
M. Azuma, Z. Hiroi, M. Takano, K. Ishida and Y. Kitaoka,
Phys. Rev. Lett. {\bf 73}, 3463 (1994).

\bibitem{batlogg_et_al95}
B. Batlogg et al., Bull. Am. Phys. Soc. 40, 327 (1995).

\bibitem{ferro_experiment}
Y. Hosokoshi et al.,
Phys. Rev. {\bf B} 60, 12924-12932 (1999).

\bibitem{greiner}
M. Greiner, O. Mandel, T. Esslinger, Th. W. H\"ansch, and I.
Bloch, Nature \textbf{415}, 39 (2002); M. Greiner, O. Mandel, Th. W.
H\"ansch, and I. Bloch, Nature \textbf{419}, 51 (2002)


\bibitem{OLs}
J. J. Garcia-Ripoll, M. A. Martin-Delgado, J. I. Cirac,
"Implementation of Spin Hamiltonians in Optical Lattices";
Phys. Rev. Lett. {\bf 93}, 250405 (2004); cond-mat/0404566.




\bibitem{white92}
S.R. White, 
{\em ``Density-matrix algorithms for quantum renormalization groups''};
Phys. Rev. Lett.{\bf 69}, 2863 (1992).

\bibitem{white93}
S.R. White, Phys. Rev. B \textbf{48}, 10345 (1993)

\bibitem{hallberg06}
K. Hallberg,
"New Trends in Density Matrix Renormalization";
Adv.Phys. {\bf 55}  477-526 (2006); cond-mat/0609039.


\bibitem{scholl05}
A. Schollw\"{o}ck,
\textit{The density matrix renormalization group},
Rev. Mod. Phys.,{\textbf 77}, 259 (2005).

\bibitem{dmrg_book}
\textit{Density Matrix Renormalization}, edited by I.
Peschel, X. Wang, M. Kaulke and K. Hallberg (Series: Lecture Notes in
Physics, Springer, Berlin, 1999)




\bibitem{sop1} M. den Nijs, K. Rommelse,
``Preroughening transitions in crystal surfaces and valence-bond phases in quantum spin 
chains'',
Phys. Rev. B {\bf 40}, 4709 (1989).

\bibitem{sop2}
T. Kennedy, H. Tasaki,
``Hidden $Z_2{\times}Z_2$ symmetry breaking in Haldane-gap antiferromagnets'',
Phys. Rev. B {\bf 45}, 304 (1992).

\bibitem{oshikawa92}
M. Oshikawa,
"Hidden $Z_2 \times Z_2$ symmetry in quantum spin chains with
arbitrary integer spin";
3. Phys.: Condens. Matter {\bf 4} 7469-7488 (1992).



\bibitem{TOP_Nakamura_Todo}
M. Nakamura, S. Todo, ``Order parameter to characterize Valence-Bond solid states
in quantum spin chains''. Phys. Rev. Lett. \textbf{89}, 77204 (2002). 


\bibitem{TOP_Nakamura_Todo2}
M. Nakamura, S. Todo, ``Novel Order Parameter to Characterize Valence-Bond-Soli\
d States''.
Prog.  Theor. Phys. Supplement, {\bf 145}, 217-220 (2002);
arXiv:cond-mat/0201204.

\bibitem{matsumoto_et_al03}
M. Matsumoto,T. Sakai, M. Sato, H. Takayama, S.  Todo,
``Quantum phase transitions of spin chiral nanotubes'';
Physica {\bf E} 29, 660 (2005);
arXiv:cond-mat/0506626).

\bibitem{LSM}
E. Lieb, T. Schultz, and D. Mattis,
``Two soluble models of an antiferromagnetic chain'',
Ann. Phys. (N.Y.) {\bf 16}, 407-466 (1961).

\bibitem{affleck_lieb86}
I. Affleck and E.H. Lieb,
``A proof of part of Haldane's conjecture on quantum spin chains'',
Lett. Math. Phys. {\bf 12}, 57-69 (1986).

\bibitem{affleck88}
I. Affleck,
``Spin gap and symmetry breaking in CuO$_2$ layers and other antiferromagnets'',
Phys. Rev. {\bf B} 37, 5186 - 5192 (1988).


\end{thebibliography}
\end{document}